\documentclass[aps,prb,twocolumn,showpacs]{revtex4}
\usepackage{graphics,amsmath}
\usepackage{tabularx,graphicx}
\usepackage[latin1]{inputenc}
\usepackage{epsfig}

\begin{document}

\title{Variational approach to the excitonic phase transition in graphene}
\author{J. Sabio$^{1,2}$}
\author{F. Sols$^2$}
\author{F. Guinea$^1$}
\affiliation{$^1$Instituto de Ciencia de Materiales de Madrid
(CSIC), Sor Juana In\'es de
la Cruz 3, E-28049 Madrid, Spain. \\
$^2$ Departamento de F{\'\i}sica de Materiales, Universidad
Complutense de Madrid, 28040 Madrid, Spain.}

\begin{abstract}
We analyze the Coulomb interacting problem in undoped graphene
layers by using an excitonic variational ansatz. By minimizing the
energy, we derive a gap equation which reproduces and extends known
results. We show that a full treatment of the exchange term, which
includes the renormalization of the Fermi velocity, tends to suppress the phase transition by increasing the critical coupling at which the excitonic instability takes place.
\end{abstract}

 \maketitle

{\em Introduction.}
The role of Coulomb interactions in the low-energy regime of undoped
graphene layers has arisen great interest and still remains
somewhat controversial. This is rooted to the poor screening
properties of graphene, a peculiarity that can be traced back to the
linear density of states of the low-energy theory, which vanishes at
the Dirac point. An early weak-coupling analysis of the problem,
based on the Renormalization Group (RG) method, showed that the
Coulomb interaction is marginally irrelevant, flowing to a
non-interacting fixed point.\cite{Gon94} This picture can be
rigorously justified at all couplings in the limit of a large number
of electron flavors.\cite{Gon99} In that scenario, undoped graphene
layers would behave mostly as a non-interacting system of electrons,
with minor traces of interactions reflected in the lifetime of
quasiparticles \cite{Gon96} and in a logarithmic renormalization of the
Fermi velocity.\cite{Gon94} Remarkably,
this picture seems to match reasonably well with current experimental data.%
\cite{Li08, Nair08, Li09}

The relative strength of the Coulomb interaction measured, as
compared with the kinetic energy of the electrons, is ruled by the
dimensionless coupling
constant $g_0\equiv e^{2}/\epsilon v_{F}$, where $e$ is the electron charge, $%
v_{F}$ the Fermi velocity, and $\epsilon $ the dielectric constant
of the medium in which graphene is embedded. We use the subscript $0$ to denote unscreened values of the coupling, see below. For samples in
vacuum, $e^2 \simeq 14.4$ eV \AA, and $v_F \simeq 6.3$ eV \AA, so that $g_0\simeq 2.3$. Density Functional Theory\cite{Polini08y2,SK10} give a value for the screened coupling in the range
$g \sim 0.5-2$.  This puts graphene in the
intermediate coupling regime, and hence the validity of the
weak-coupling analysis relies on the absence of a strong coupling
fixed point in the RG transformation. Indications of such a fixed
point have been found by extending the weak-coupling RG to higher
orders in the coupling constant expansion. \cite{Vaf08} The
experimental data would still be compatible with this strong-coupling
scenario if current setups had graphene sufficiently
isolated from the environment and could operate with perfectly neutral
samples.

The possibility of phases beyond the reach of perturbative or weak
coupling renormalization group methods in undoped graphene has been
explored in the literature by using different approaches. The main
candidate for a strong coupling phase is an excitonic condensate, in
which electron and holes bind together opening up a gap in the
density of states and rendering the system insulating. The mechanism
responsible for this phase would be the gain in exchange energy arising
from the long range Coulomb interaction. A gap equation for
this transition has been derived within the Dyson-Schwinger formalism, \cite%
{K01} and different solutions of this equation yield (unscreened) critical
couplings for the phase transition around an unscreened coupling $g_{0c}\sim 1-2$.
\cite{K01, K09, Gamayun10} This scheme is equivalent to the
summation of a class of diagrams, and can be considered an extension
of weak coupling approaches. Montecarlo calculations in the lattice
have been carried out to analyze this
problem, \cite{Drut09, Armour10} finding an insulating phase above $%
g_{0c}\simeq 1.11$ and $g_{0c}\simeq 1.66$, respectively. A phase
transition beyond a certain coupling can also be found for short
range interactions in the half filled honeycomb lattice,\cite{ST92}
although the critical coupling takes the model beyond the regime
where the approximation of the electronic bands by the Dirac
equation is valid. Finally, the study of the two-body problem
in graphene, with Coulomb interactions, leads to a
remarkable instability of the wave-function for a critical coupling $g_{0c}=1$%
,\cite{Sabio09} that might underlie the eventual formation of
excitons.

In this paper we investigate the possibility of an excitonic strong
coupling phase in undoped graphene by using a variational ansatz.
The method used here can be extended in a straightforward
way to finite temperatures or to finite carrier concentrations. We
derive a
gap equation similar to that obtained in the previous literature, \cite%
{K01, Gamayun10} but with the inclusion of a the renormalization of
the Fermi velocity. By analyzing numerically and analytically the resulting gap equation, we find that the latter produces a suppresion of the phase transition by increasing the critical coupling.

{\em The model and variational ansatz.}
As we have mentioned, we will address the problem of an undoped
graphene sample with Coulomb interactions in the low-energy regime,
where the electron motion is described by the Dirac equation. The
Hamiltonian for this problem reads:
\begin{equation}
{\mathcal{H}}=\sum_{\mathbf{k}s}skn_{\mathbf{k}s}+\frac{1}{2}\sum_{\mathbf{q}%
}V_{q}n_{\mathbf{q}}n_{-\mathbf{q}}
\end{equation}%
where $s=\pm $ refers to the upper and lower cones, respectively. In
our model, the spin and valley degrees of freedom are considered
only as extra
degeneracies in the number of fermions. The Coulomb potential is given by $%
V_{q}=2\pi g/q$, where $g$ is the dimensionless coupling constant
introduced above. We assume that $g$ includes contributions from static screening\cite{Gon99,K01}. In the RPA approximation, transitions between the valence and conduction bands lead to a momentum independent dielectric constant, which can be incorporated in a straightforward into this formalism. Alternatively, one can view this approach as the leading approximation in the limit $N \rightarrow \infty$, where $N$ is the number of fermion flavors\cite{Fos08}.

Our goal is to analyze the ground state of this Hamiltonian by
using a variational ansatz which includes the possibility of pairing
between electron and holes. Such an ansatz was proposed to study
excitons formation in semiconductors,\cite{Comte82} and is
reminiscent of the ansatz used in the BCS theory of
superconductivity:
\begin{equation}
|\Psi \rangle =\Pi _{\mathbf{k}}(u_{k}+v_{k}c_{\mathbf{k}%
+}^{\dagger }c_{\mathbf{k}-})|D\rangle
\end{equation}%
The ansatz contains a coherent superposition of states with a
different number of electron-hole pairs. Here $|D\rangle $ stands
for the filled Dirac sea, and $u_{k}$ and $v_{k}$
are variational parameters to be determined by minimizing the ground
state energy. Without loss of generality, they are taken real.
Notice that they are not independent, since the normalization of the
wave function imposes the constraint:
\begin{equation}
u_{k}^{2}+v_{k}^{2}=1
\end{equation}

\begin{figure}[bt]
\begin{center}
\includegraphics[width=0.9\columnwidth]{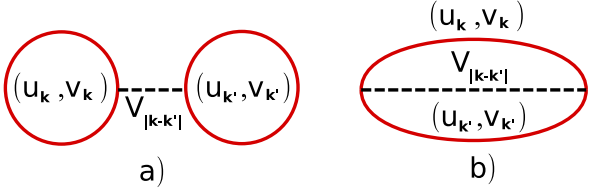}
\end{center}
\caption{Diagrams included in the ground state energy within the
variational ansatz. a) Hartree term, which is zero. b)
Exchange term, which is the dominant one. } \label{fig1}
\end{figure}

{\em Derivation of the gap equation.} Following the lines of a typical variational calculation, the energy
of the ansatz is evaluated by projecting the Hamiltonian into this
state. It has
two contributions, the Hartree and the exchange one, as shown in Fig. \ref%
{fig1}. The Hartree contribution is zero, by virtue of the normal
ordering of the Hamiltonian with respect to the Dirac sea, which is
physically related to the neutrality of charge of the global system.
The dominant contribution comes from the exchange energy, which
includes terms with a momentum transfer of
$\mathbf{q}=\mathbf{k}^{\prime }-\mathbf{k}$. The projected
Hamiltonian reads then:
\begin{eqnarray}
\langle \Psi | :{\mathcal H}: |\Psi \rangle = \sum_{{\bf k}}  k
(v_{k}^2 - u_{k}^2)  \nonumber  \\ - \frac{1}{2} \sum_{{\bf
k},{\bf k}'} V_{|{\bf k}'-{\bf k}|} \left[ 2 u_k u_{k'} v_k v_{k'} \right. \nonumber \\ \left. +
\cos^2\left(\frac{\theta_{\bf k'} - \theta_{\bf k}}{2}\right)(u_k^2
u_{k'}^2 + v_k^2 v_{k'}^2)  \nonumber \right. \\ \left. + \sin^2\left(\frac{\theta_{\bf k'} -
\theta_{\bf k}}{2}\right)(u_k^2 v_{k'}^2 + v_k^2 u_{k'}^2) \right]
\end{eqnarray}
where we have used the normal-ordered Hamiltonian in order to carry out the calculation. The extreme condition must be imposed respecting the
normalization constraint. The result
gives the following equation:
\begin{eqnarray}
&&\left[ k+\sum_{\mathbf{k}^{\prime }}V_{|\mathbf{k}^{\prime }-\mathbf{k}%
|}\cos (\theta _{\mathbf{k^{\prime }}}-\theta _{\mathbf{k}})(u_{k\mathbf{%
^{\prime }}}^{2}-v_{k\mathbf{^{\prime }}}^{2})\right] u_{k}v_{k}  \notag \\
&=&(u_{k}^{2}-v_{k}^{2})\sum_{\mathbf{k}^{\prime }}V_{|\mathbf{k}^{\prime }-%
\mathbf{k}|}u_{k\mathbf{^{\prime }}}v_{k\mathbf{^{\prime }}}
\end{eqnarray}%
This equation can be simplified by introducing the following
parameters:
\begin{eqnarray}
\xi _{k} &=&k+\sum_{\mathbf{k}^{\prime }}V_{|\mathbf{k}^{\prime }-\mathbf{k}%
|}\cos (\theta _{\mathbf{k}^{\prime }}-\theta
_{\mathbf{k}})(u_{k^{\prime
}}^{2}-v_{k^{\prime }}^{2})  \label{linear} \\
\Delta _{k} &=&2\sum_{\mathbf{k}^{\prime }}V_{|\mathbf{k}^{\prime }-\mathbf{k%
}|}u_{k^{\prime }}v_{k^{\prime }}  \label{gap0} \\
E_{k}^{2} &=&\xi _{k}^{2}+\Delta _{k}^{2}
\end{eqnarray}%
The first equation is the self-energy insertion to the electron
propagator, which adds to the linear term coming from the
non-interacting dispersion relation and represents a renormalization
of the Fermi velocity. The second equation introduces $\Delta _{k}$,
which can be identified with the gap that arises in the electronic
spectrum when excitons are formed. This is clearly expressed in the
third equation, which gives the dispersion relation of Bogoliubov
quasiparticles in the excitonic condensante.

In terms of these new parameters, the solution to the variational
problem reads:
\begin{eqnarray}
u_{k}v_{k} &=&\frac{\Delta _{k}}{2E_{k}} \\
v_{k}^{2} &=&\frac{1-\xi _{k}/E_{k}}{2} \\
u_{k}^{2} &=&\frac{1+\xi _{k}/E_{k}}{2}
\end{eqnarray}%
By plugging these expressions into the equation for the gap, Eq. (\ref{gap0}%
), we get a self-consistent integral equation, namely:
\begin{equation}
\Delta _{k}=\sum_{\mathbf{k^{\prime }}}V_{|\mathbf{k}^{\prime }-\mathbf{k}|}%
\frac{\Delta _{k\mathbf{^{\prime }}}}{E_{k\mathbf{^{\prime }}}}
\end{equation}%
As we have already mentioned, a similar gap equation has been
already found by using the Schwinger-Dyson formalism. \cite{K01}

Further insight can be obtained by carrying out the angular integral
while keeping the lowest order terms in a Legendre polynomial
expansion of the Coulomb interaction $V_{|\mathbf{k}^{\prime
}-\mathbf{k}|}$. This yields a
simplified integral equation in the continuum limit of the problem:%
\begin{equation}
\Delta _{k}=g\int_{0}^{\Lambda }dk^{\prime }k^{\prime }\Delta _{k^{\prime }}%
\frac{{\mathcal{K}}(k,k^{\prime })}{\sqrt{\xi _{k^{\prime
}}^{2}+\Delta _{k^{\prime }}^{2}}}~,
\end{equation}%
where we have introduced the following kernel:
\begin{equation}
{\mathcal{K}}(k,k^{\prime })=\frac{1}{k}\theta (k-k^{\prime })+\frac{1}{%
k^{\prime }}\theta (k^{\prime }-k)~.
\end{equation}%
The main feature of this gap equation, as compared with previous
approaches, is the inclusion of the exchange correction to the free
electron dispersion relation, Eq. (\ref{linear}).

{\em Analysis of the gap equation.}
In order to extract information from the gap equation, we make the
following assumption:\cite{K01,Gorbar02} the dominant contribution
to the gap equation corresponds to the region $k\gg \Delta ^{\ast
}$, where $\Delta ^{\ast }\equiv \Delta _{\Delta ^{\ast }}$ (with
$v_{F}\equiv 1$). This allows us to make $E_{q}\simeq \xi _{q}$ and
write $\Delta ^{\ast }$ as the lower limit of the integral:
\begin{equation}
\Delta _{k}\simeq g\int_{\Delta ^{\ast }}^{\Lambda }dk^{\prime
}k^{\prime }\Delta _{k^{\prime }}\frac{{\mathcal{K}}(k,k^{\prime
})}{\xi _{k^{\prime }}} \label{integral}
\end{equation}%
By using the same type of reasoning, an expression for the $\xi
_{k}$ can be derived, which only retains the leading, most divergent
terms (and valid for $\xi _{k}>\Delta^{*}$):
\begin{equation}
\xi _{k}=k+\frac{g}{4}k\log \left( \frac{\Lambda }{k}\right)
\end{equation}%
As mentioned above, this is actually the renormalization of the
Fermi velocity that arises from a RG analysis, \cite{Gon94} which
has been so far neglected in the literature on the excitonic
condensation. We will see shortly that this logarithmic correction
plays a crucial role in the analysis of the gap equation.

Let us transform the integral equation into a differential equation,
namely:
\begin{equation}
k^{2}\Delta _{k}^{\prime \prime }+2k\Delta _{k}^{\prime }+g(k)\Delta
_{k}=0 \label{differential}
\end{equation}%
which has the form of a 3D radial Schr\"odinger equation with a
potential $g(k) $. The latter is the running coupling constant in
the RG sense, which has appeared in a natural way from the exchange
correction to the linear dispersion relation. It reads:
\begin{equation}
g(k)=\frac{g}{1+\frac{g}{4}\log \left( \frac{\Lambda }{k}\right) }
\label{renormalized}
\end{equation}%
The differential equation (\ref{differential}) must be supplemented
with boundary conditions that are also derived from Eq.
(\ref{integral}):
\begin{eqnarray}
k^{2}\Delta _{k}^{\prime }|_{k=\Delta ^{\ast }} &=&0 \label{boundary1}\\
(k\Delta _{k}^{\prime }+\Delta _{k})|_{k=\Lambda } &=&0
\label{boundary2}
\end{eqnarray}%
The first one is the infrared condition, since it is evaluated at the gap $%
\Delta ^{\ast }\ll \Lambda$, while the second one is the ultraviolet one,
evaluated at the cutoff.

{\em Adiabatic solution.}
A preliminary study of Eqs. (\ref{differential})-(\ref{boundary2}) can be made by assuming that $g(k)$ varies
slowly enough for an adiabatic approximation to be reasonable. Noting that
the case of a constant potential $g(k)=g$ admits an exact solution of the form%
\[
\Delta _{k}=Ak^{-\frac{1}{2}(1+\sqrt{1-4g})}+Bk^{-\frac{1}{2}(1-\sqrt{1-4g}%
)}~,
\]%
the adiabatic solution can be found to be, following Ref. \onlinecite{K09},%
\begin{eqnarray}
\Delta _{k}^{\text{ad}}&=&\frac{C_{+}e^{i\varphi(k)}+C_{-}e^{-i\varphi(k)}}{\sqrt{k}\left[ g(k)-\small{\frac{1}{4}}\right]^{1/4}}
~,
\label{adiabatic}
\end{eqnarray}
where $\varphi(k)\equiv \int_{\Delta ^{\ast }}^{k}\frac{dk^{\prime }}{k^{\prime }%
}\sqrt{g(k^{\prime })-\small{\frac{1}{4}}}~.$
Implementation of the boundary conditions (\ref{boundary1})-(\ref{boundary2}) yields the quantization rule%
\begin{equation}
\varphi(\Lambda)%
+\delta _{\Lambda }+\delta _{\Delta ^{\ast }}=\pi n~,
\label{quantization}
\end{equation}%
where $n$  is a positive integer and $\delta _{k}\equiv \arctan \sqrt{4g(k)-1%
}$. The goal is to solve for $\Delta ^{\ast }$, a nonzero value meaning that
there is an excitonic instability. The condition $g(k)>1/4$ for all values
of $k$, leads to the requirement $\Delta ^{\ast }>\Delta _{\min }=\Lambda e^{-8(1-1/4g)}$. We find that a nonzero, real solution of Eq. (\ref{quantization}) (with $n=1$) satisfying $\Delta ^{\ast }>\Delta _{\min }$, exists for $g$ greater than a critical value $g_c\simeq 0.5$, which marks the onset of the excitonic instability.

\begin{figure}[bt]
\begin{center}
\includegraphics[width=0.45\columnwidth]{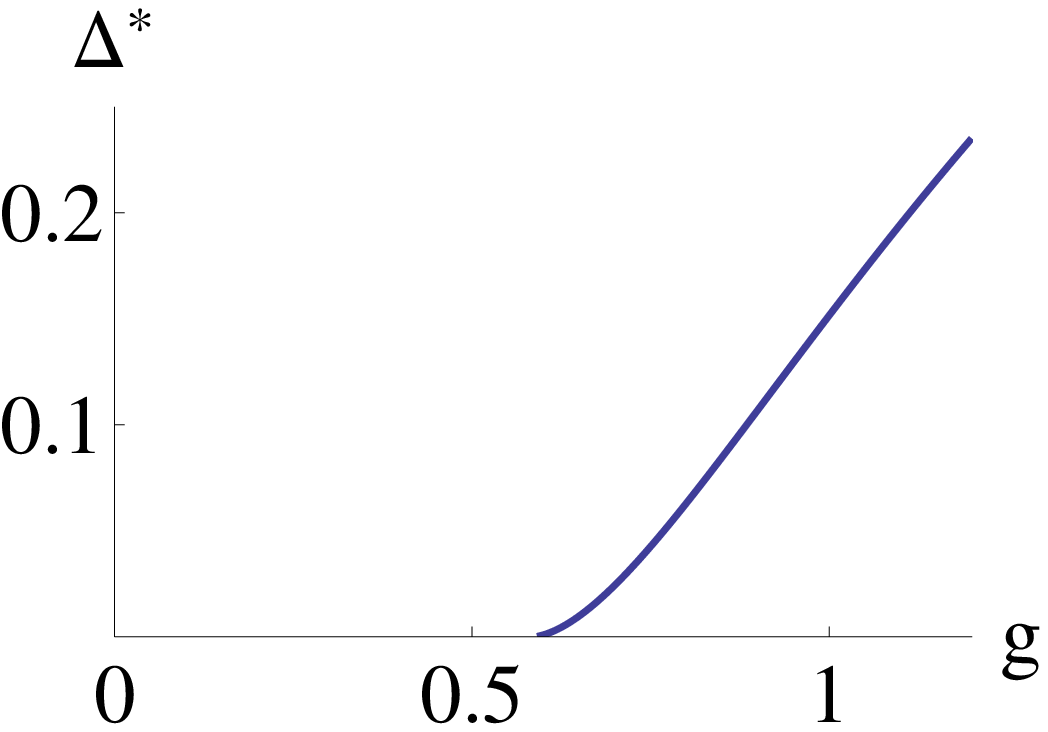}
\includegraphics[width=0.45\columnwidth]{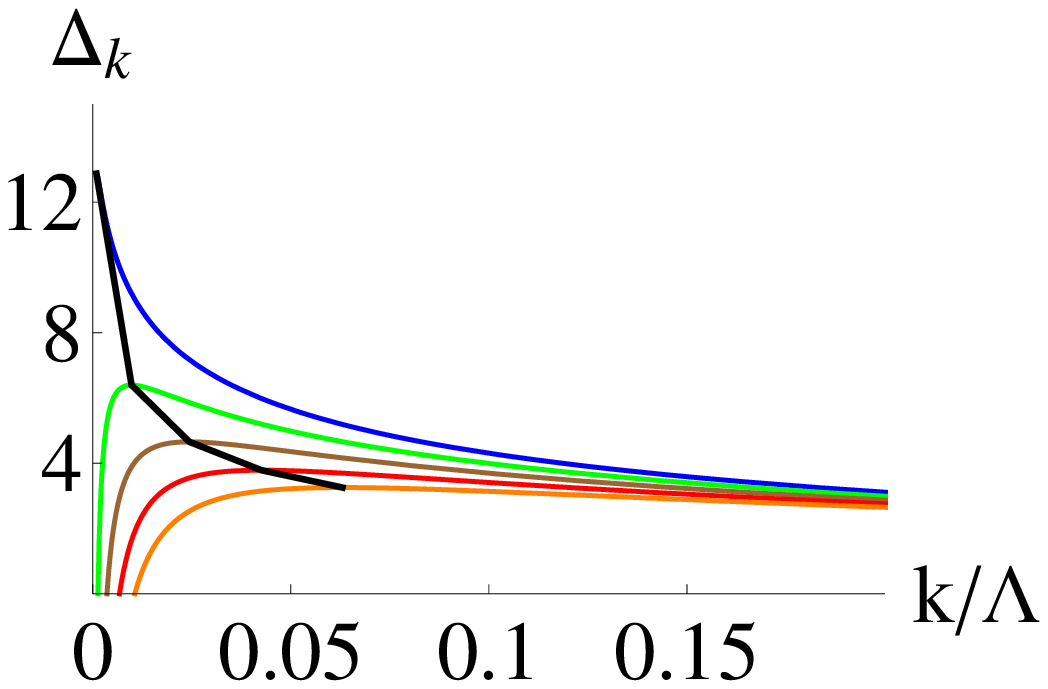}
\end{center}
\caption{(Color online). Left: Dependence of  $\Delta^*$ on $g$ obtained by numerically integrating Eq.~(\ref{differential})
with the boundary conditions in Eqs.~(\ref{boundary1})-(\ref{boundary2}). Right: Dependence of $\Delta_k$ on $k$ for $g = 0.6 , 0.65 , 0.7 , 0.75$ and $0.8$ (from top to bottom). The thick black line shows the
position of the maxima of $\Delta_k$, which give the value of $\Delta^*$, see Eq.~(\ref{boundary1}).} \label{fig_delta}
\end{figure}

\begin{figure}[bt]
\begin{center}
\includegraphics[width=0.45\columnwidth]{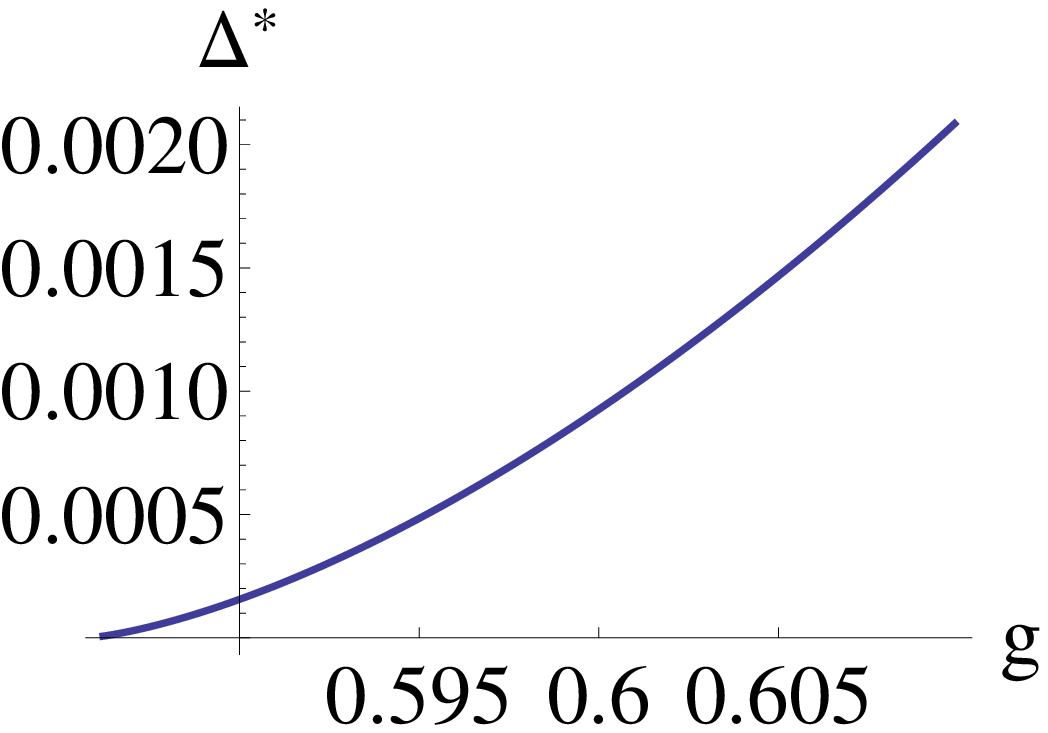}
\includegraphics[width=0.45\columnwidth]{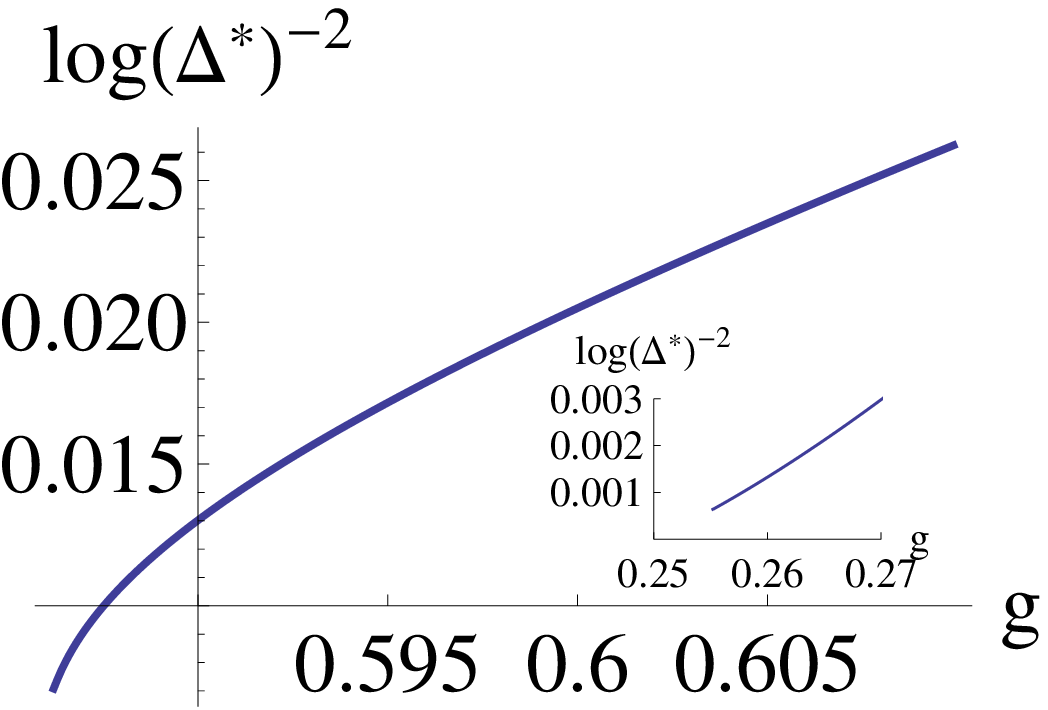}
\end{center}
\caption{(Color online). Details of the dependence of $\Delta^*$ on $g$ near the transition. The inset on the right graph shows the results from numerically solving Eq.~(\ref{differential}) without velocity renormalization [i.e. assuming $g(k)=g$ in Eq. (\ref{renormalized})].} \label{fig_gc}
\end{figure}

{\em Numerical solution.}
We further check the previous analysis by numerically solving Eq.~(\ref{differential}) with the boundary conditions (\ref{boundary1})-(\ref{boundary2}). The results are
shown in Fig.~\ref{fig_delta}. We find solutions for $g \ge g_c \simeq 0.59$, in reasonable agreement with the adiabatic approximation. The asymptotic limit $\Delta_k
\sim 1 / \sqrt{k}$ [see Eq. (\ref{adiabatic})] is only clearly visible for $g \sim g_c$ and $k \ll \Lambda$. A detailed analysis of the region where $(g -g_c)\rightarrow 0^{+}$ suggests that $\Delta^{\ast } \propto ( g - g_c )$, see Fig.~\ref{fig_gc}. For comparison, we also show the numerical results obtained by neglecting the renormalization of the Fermi velocity. They reproduce correctly the main features found in analytical studies, namely, $g_c = 1/4$ and $\Delta^* \sim e^{- A / \sqrt{g-g_c}}$, where $A$ is a constant.\cite{K01,K09, Gamayun10}

{\em Conclusions.}
We have analyzed the problem of Coulomb interactions in undoped
graphene by using a variational ansatz that includes the possibility
of exciton formation. Our approach can be readily extended to other two- and three-dimensional materials, as well as to finite temperatures and carrier concentrations.\cite{HRV10} It allows us to calculate the total free energy, which can be compared to that of other broken symmetry phases.

Our variational analysis reproduces the main features of the excitonic transition in graphene.\cite{K01}
In addition, we find that a renormalization of the Fermi
velocity is a natural by-product of the variational treatment. The resulting change in one particle energies leads to a cancelation of
the leading divergences with trigger the excitonic transition. A similar effect is observed in the analysis of the excitonic transition
due to short range interactions in graphene bilayers.\cite{NL10,LATF10} Our variational analysis leads to a critical coupling $g_c \simeq 0.59$, which is about a factor
two larger than the critical coupling obtained neglecting the Fermi velocity renormalization, $g_c = 1/4$.\cite{K01,GGMS02,Khve08,Gamayun10} The renormalization of $g$ to two loops gives a transition at $g_c \simeq 0.83$.\cite{Vaf08}

In suspended graphene, we can assume that the effective value of $g$
is modified solely by internal screening,\cite{Gon99,K01} so that $g = g_0 / ( 1 + N \pi g_0 /8)$, where $N=4$ is the number of electron flavors, and $g_0 = e^2 / v_F$
is the bare coupling constant. Then, an upper bound of $g$ is
$g_{\rm max} = \lim_{e^2 / v_F \rightarrow \infty} g ( e^2 / v_F ) = 2 / \pi \simeq 0.64$, which lies slightly above the value of $g_c$ obtained with our variational ansatz. For the realistic value of $g_0=2.3$ we obtain $g=0.50$, which is below the critical value $g_c\simeq 0.59$ which
we have found by solving Eq. (\ref{differential}) numerically.\cite{montecarlo}

{\em Acknowledgements.}
We acknowledge financial support by MICINN (Spain) through grants
FIS2007-65723, FIS2008-00124 and CONSOLIDER CSD2007-00010, and by
the Comunidad de Madrid, through NANOBIOMAG and MICROSERES. We also acknowledge
useful discussions with J. Gonz\'alez, M. I. Katsnelson, M. Polini,
and I. Zapata.

\bibliography{excitonic}

\begin{thebibliography}{25}
\bibitem[{mon()}]{montecarlo}
\bibinfo{note}{Montecarlo results suggest a critical coupling in the range of
  expected values in suspended samples\cite{Drut09}, which is smaller than the
  value predicted by our variational ansatz. The origin of this discrepancy is
  unclear. It could arise from differences in the regularization of the short
  range part of the interaction, or in the treatment of screening at high
  frequencies.}
\end{thebibliography}

\end{document}